\documentclass[jpcck,article]{achemso}
\usepackage{graphicx} 
\usepackage{chngcntr}

\author{M. Nandi}
\author{N. Khan}
\author{D. Bhoi}
\author{A. Midya}
\author{P. Mandal}
\email{prabhat.mandal@saha.ac.in}
\affiliation{Saha Institute of Nuclear Physics, 1/AF Bidhannagar, Calcutta 700 064, India}
\title{Field-Induced Spin-Structural Transition and Giant Magnetostriction in Ising Chain $\alpha$-CoV$_2$O$_6$}

\begin{document}
\begin{abstract}

We have investigated the temperature and magnetic field dependence of magnetization, specific heat ($C_p$), and  relative sample length change ($\Delta L/L_0$) for understanding the field-induced spin-structural change in quasi-one-dimensional spin chain $\alpha$-CoV$_2$O$_6$ which undergoes antiferromagnetic (AFM) transition below $T_N$$=$15 K. Analysis of $C_p$($T$) shows that an effective $S$$=$1/2 Ising state is realized below 20 K, though the magnetic fluctuations persist well above $T_N$. $C_p$ and the coefficient of linear thermal expansion ($\alpha$) exhibit strong  $H$ dependence in the AFM  state. We also observe a huge positive magnetostriction [$\Delta L$($H$)/$L_0$] below 20 K which does not show any tendency of saturation up to 9 T. With increasing field, a sharp and symmetric peak emerges  below $T_N$ in both $C_p$($T$) and  $\alpha$($T$) due to  field-induced first order ferrimagnetic/ferromagnetic-paramagnetic transitions. The large value of magnetostriction  below $T_N$ suggests strong spin-lattice coupling in $\alpha$-CoV$_2$O$_6$. \\

\end{abstract}
\newpage
\maketitle
\section{Introduction}
Several  cobalt-based low-dimensional  compounds and rare-earth-based pyrochlore oxides exhibit fascinating  magnetic  properties such as magnetic field ($H$) induced spin order-disorder transition \cite{he}, 1/3 magnetization plateau in the $M$($H$) curve \cite{maar},  quantum phase transition \cite{cold} and spin-structural change \cite{bram,taba,ram,hir,kimu}. Geometrical frustration due to the triangular or tetrahedral arrangement of the magnetic moments, bond frustration  as a result of competing ferromagnetic (FM) and antiferromagnetic (AFM) exchange interactions and large single ion anisotropy are the fundamental ingredients that eventually determine the complexity of the magnetic ground state and hence the  new functionalities in these compounds. Often, the ground state of the frustrated materials is extremely sensitive to external perturbations such as magnetic field.  Though there are a number of systems showing such kind of magnetic ground state, the underlying mechanism responsible for the above features has not been well understood yet.  \\

Recently,  the quasi-one-dimensional (1D) spin-chain CoV$_2$O$_6$ has received attention to the scientific community due to the observation of  1/3 magnetization plateau similar to that observed in a regular triangular lattice \cite{zhe,lene,kimb,lener,lenert,ksingh,mark,markk,yao,kim,saul}. In monoclinic $\alpha$-CoV$_2$O$_6$ and  triclinic $\gamma$-CoV$_2$O$_6$, the edge-shared CoO$_6$ octahedra form a magnetic chain along the $b$ axis, and the edge-shared VO$_5$ square pyramids are located in between the magnetic chains.  The much larger Co-Co interchain distance as compared to the intrachain Co-Co distance and the presence of the nonmagnetic V$^{5+}$ ion in between the chains weaken the interchain magnetic coupling considerably.  Both $\alpha$-CoV$_2$O$_6$ and $\gamma$-CoV$_2$O$_6$ show large single ion anisotropy, undergo long-range AFM transition below 15 and 6 K, respectively and exhibit field-induced metamagnetic transitions at two critical fields $H_{c1}$ and $H_{c2}$ \cite{zhe,lene,kimb,lener,lenert,ksingh,mark,markk}.  The  values of magnetic moment determined from the saturation magnetization ($M_S$) in the field-induced FM state and susceptibility in the paramagnetic (PM) state  are found to be significantly larger than  the expected spin-only moment of high spin-state Co$^{2+}$ ion \cite{zhe,lene,kimb}. In $\alpha$-CoV$_2$O$_6$, $M_S$ is as much as 1.5 $\mu_B$/Co larger than the spin-only moment (3.0 $\mu_B$/Co) for $H$ parallel to $c$ axis. This excess 1.5 $\mu_B$/Co moment is thought to come from the orbital magnetic moment due to the strong  spin-orbit or spin-lattice coupling \cite{zhe,lene,kimb}.\\

Magnetic, electric, and structural properties have been studied extensively to disclose the underlying mechanism responsible for the step-like jumps in $M$($H$) of CoV$_2$O$_6$ \cite{zhe,lene,kimb,lenert,ksingh,mark,markk,lener}.  Neutron diffraction studies have revealed that  these jumps in $M$($H$) curve are coupled  to the field-induced magnetic phase transitions from AFM to FM state through an intermediate ferrimagnetic state. \cite{mark,markk,lener}.  To elucidate the nature of field-induced metamagnetic  transition and the thermodynamic properties of frustrated systems, the measurement of specific heat ($C_p$) in applied field is important. In this work, we present  the specific heat data of $\alpha$-CoV$_2$O$_6$ over a wide range of $H$ across the magnetic ordering at $T_N$. Strong spin-lattice coupling  has also motivated us to investigate the magnetostriction effect [$\Delta L$($H$)/$L_0$] in this system.  We observe that both $C_p$($T$) and the linear thermal expansion coefficient $\alpha$($T$) [$=$($1/L_0$)d$\Delta L$($T$)/d$T$] exhibit a single $\lambda$-like peak at $T_N$ in absence of magnetic field but a sharp and symmetric peak appears below $T_N$ with applied field. Also, the system releases the Ising-like entropy (Rln2) and exhibits very large magnetostriction effect below $T$$\sim$ 20 K. For understanding the origin of the field-induced magnetic transition in $\alpha$-CoV$_2$O$_6$, the present results have been  compared and contrasted with well known frustrated systems where the  field-induced metamagnetic  transition  occurs due to the triangular or tetrahedral crystal geometry of the magnetic moments. \\

\section{Experimental details}
Polycrystalline  $\alpha$-CoV$_{2}$O$_{6}$ samples were prepared by standard solid-state reaction method using high purity cobalt (II) acetate tetrahydrate (Aldrich, 99.999$\%$) and vanadium pentoxide (Alfa Aesar, 99.995$\%$). Stoichiometric quantities of these compounds were mixed properly  and  the mixture was heated in air for 16 h at 650 $^o$C and then at 725 $^o$C for 48 h. After the heat treatment, the material was quenched in liquid nitrogen  to obtain single phase $\alpha$-CoV$_{2}$O$_{6}$. Phase purity was checked by powder x-ray diffraction (XRD) method with CuK$_{\alpha}$ radiation in a high resolution Rigaku TTRAX II  diffractometer. We have not observed any impurity phase within the resolution ($\sim$2$\%$) of XRD. All the peaks in the diffraction pattern were fitted well to a  monoclinic structure of space group $C$2/$m$ using the Rietveld method with an acceptance factor $R_{Bragg}$($\%$)$=$3.78 (Fig. 1). The observed lattice parameters  \emph{a}=9.2501 {\AA}, \emph{b}=3.5029 {\AA}, \emph{c}=6.6175 {\AA}, and $\beta$ = $111.61^o$ are in good agreement with the reported values for $\alpha$-CoV$_{2}$O$_{6}$ \cite{lene}. We have determined the average grain size ($\sim$35$\mu$m) in the sample using the scanning electron microscope (FEI Quanta 250).  The sintered powder was pressed into pellet and then the sample of dimensions 1.5$\times$1.3$\times$1.08 mm$^3$ was cut from this pellet for the specific heat, magnetostriction and thermal expansion measurements. The longitudinal magnetothermal expansion  was measured by capacitive method using a miniature tilted-plates dilatometer with applied field parallel to thickness ($t$=1.08 mm) of the sample.  The magnetic measurements were done in both physical property measurement system and SQUID-VSM (Quantum Design). The  specific heat was measured by conventional relaxation time method using  physical property measurement system (Quantum Design). \\

\begin{figure}[h]
\includegraphics[width=0.6\textwidth]{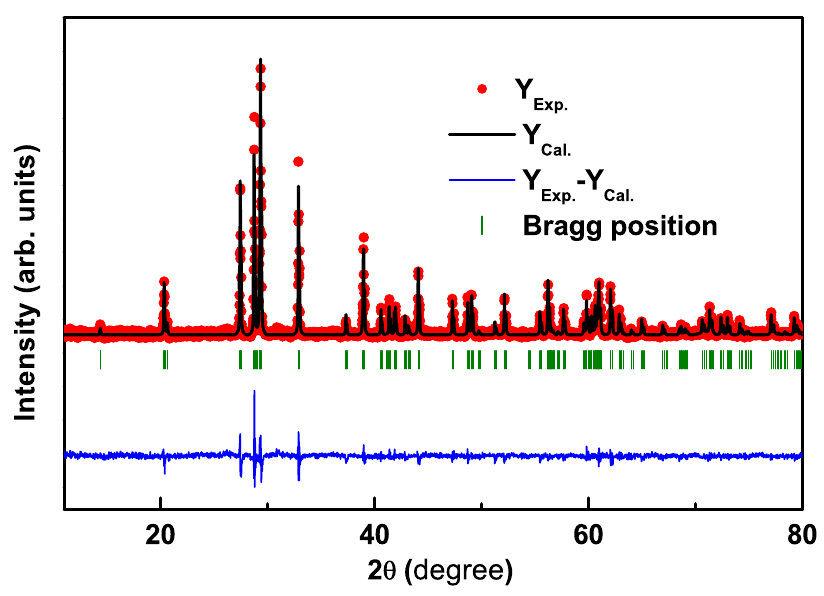}
\caption{ The x-ray powder diffraction pattern for $\alpha$-CoV$_2$O$_6$ at room temperature. The black solid line corresponds to the Rietveld refinements of the diffraction pattern.}\label{fig1}
\end{figure}

\section{Results and discussion}
Figure 2(a) displays the temperature dependence of dc magnetic susceptibility $\chi$ ($=$$M/H$) of polycrystalline  $\alpha$-CoV$_2$O$_6$ sample in a field of 0.1 T. $\chi$ increases with decreasing $T$ and shows a sharp peak at $T_N$$=$15 K due to the transition from PM to AFM state. No other peak or anomaly is observed in $\chi$($T$) in the measured temperature range (2-300 K). This indicates that the studied sample does not contain any magnetic impurity. Figure shows that $\chi^{-1}$($T$) can be fitted  well to  the Curie-Weiss law [$\chi$$=N\mu_{eff}^2/3k(T-\theta$)] over a wide range of temperature.  From this fit, we have calculated the value of effective paramagnetic moment $\mu_{eff}$$=$5.4 $\mu_B$/Co ion and the Weiss temperature $\theta$$=$-9.2 K. The negative value of $\theta$ implies that the predominant magnetic interaction is antiferromagnetic in nature. The observed value of $\mu_{eff}$ is significantly larger than  the expected spin-only moment of high spin Co$^{2+}$ (3.87 $\mu_B$). It may be mentioned that Markkula et al. have reported even larger value of $\mu_{eff}$ (6.09 $\mu_B$/Co) for this system \cite{mark}.  This huge discrepancy  between the observed and expected spin-only moment is due to the strong spin-orbit coupling \cite{zhe,lene,kimb,ksingh,mark,markk,lener}.\\
\begin{figure}[h]
\includegraphics[width=0.55\textwidth]{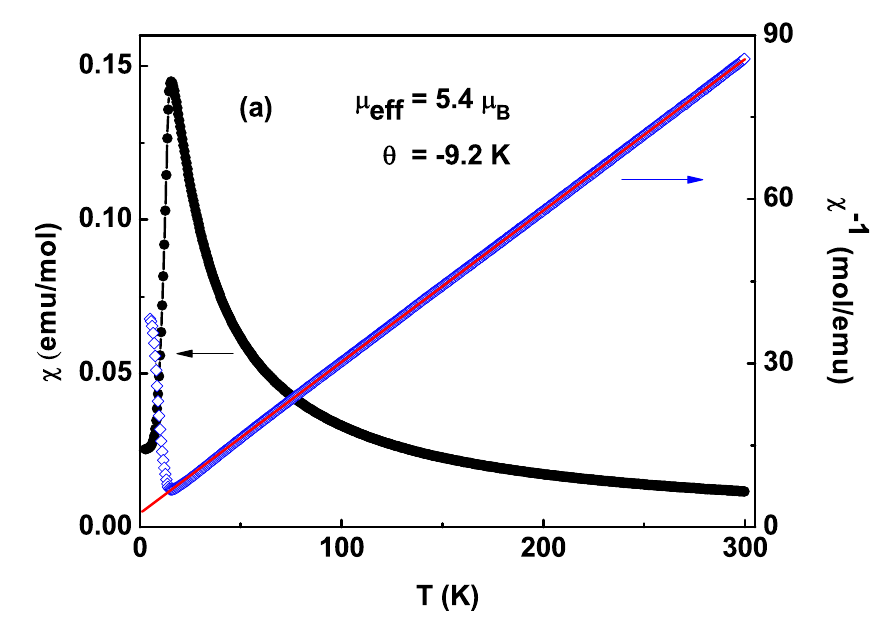}
\includegraphics[width=0.45\textwidth]{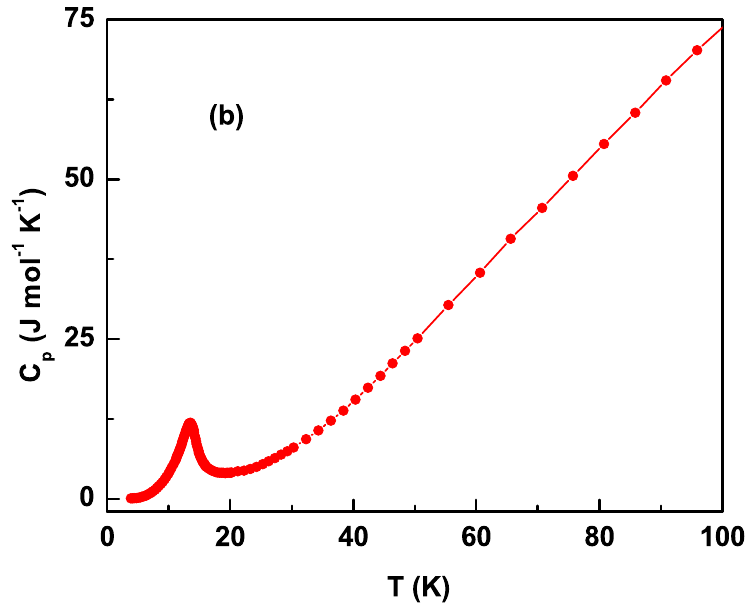}
\includegraphics[width=0.5\textwidth]{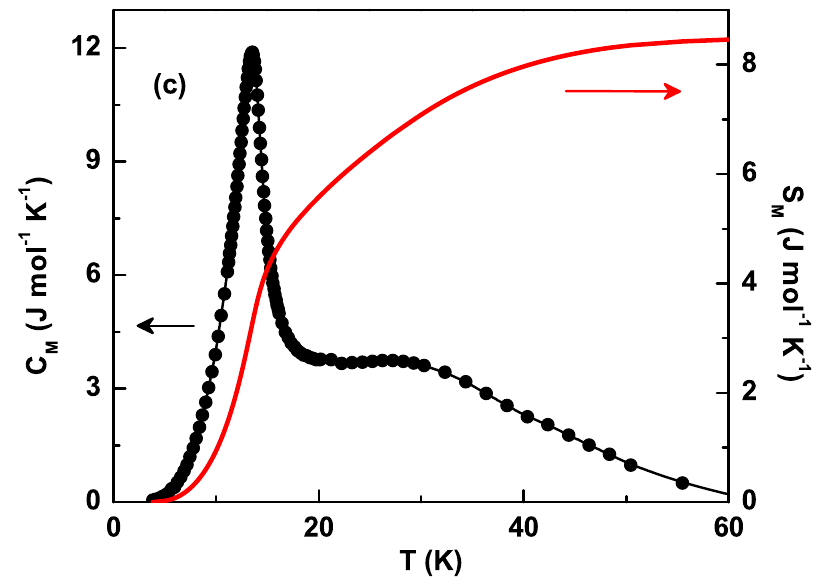}
\caption{ (a) Plots of $\chi$($T$) and $\chi^{-1}$($T$) for $\alpha$-CoV$_2$O$_6$ at 0.1 T field. Solid line is the Curie-Weiss fit to the $\chi^{-1}$($T$). (b) The temperature dependence of the specific heat $C_p$. (c) The temperature dependence of magnetic contribution to the specific heat ($C_M$) and entropy removal ($S_M$).}\label{fig2}
\end{figure}
The temperature dependence of zero-field specific heat  is shown in Fig. 2(b). With decreasing temperature, $C_p$ decreases  down to $\sim$20 K and then increases rapidly and exhibits a  $\lambda$-like anomaly close to  $T_N$. Similar to the magnetization data, $C_p$($T$) supports single magnetic phase transition in $\alpha$-CoV$_{2}$O$_{6}$. For better understanding the nature of magnetic ground state,  the magnetic contribution to the specific heat ($C_M$) in the vicinity of AFM transition and beyond has been estimated. In the temperature range 50-222 K, $C_p$ can be fitted well with the Einstein model to determine the lattice contribution to the specific heat of the system \cite{herm,lon} (please see  Figure S1 in the Supporting Information). After subtracting the lattice contribution from $C_p$, we have plotted $C_M$ in  Fig. 2(c). It is clear from Fig. 2(c) that $C_M$ does not decrease rapidly with increasing $T$ in the PM state.  The magnetic entropy ($S_M$) obtained by integrating ($C_M/T)dT$ is also shown in Fig. 2(c). $S_M$ does not saturate even at $T$ as high as 3$T_N$, which reflects the  highly anisotropic nature of the magnetic structure of $\alpha$-CoV$_2$O$_6$. We would like to mention that the observed value of  $S_M$ (=8.3 J mol$^{-1}$ K$^{-1}$ at $T$$=$3$T_N$) for the present sample is about 1.5 times larger than that reported by Kim  et al. \cite{kim}. The larger value of $S_M$ is an indication of superior ordering of spins because it means, more number of spins are participating in the magnetic transition.  The  estimated value of $S_M$,  however, is significantly smaller than the expected 11.5 J mol$^{-1}$ K$^{-1}$ for the high spin Co$^{2+}$ ($S$$=$3/2). It may be noted that $S_M$$=$5.6 J mol$^{-1}$ K$^{-1}$ at 20 K is almost in accord with the expected value  (Rln2) for the degree of freedom of the Ising moments. Thus, the entropy of the Ising-like spins is mostly released below 20 K.  As the intrachain ferromagnetic interaction in CoV$_{2}$O$_{6}$ is much stronger than the interchain AFM interaction, the intrachain spin degree of freedom is effectively frozen and the spins would behave like the Ising spins and hence, a huge reduction in magnetic entropy occurs. \\

Typical specific heat data ($C_p$/$T$) in the vicinity of $T_N$ are shown in Fig. 3 as a function of $T$ for different magnetic fields. The application of magnetic field suppresses and broadens the peak at $T_N$ and  the peak position shifts slowly towards lower temperature. Apart from these usual changes, another important feature is emerging with increasing field strength. At a field of 2 T, a weak shoulder-like feature  starts to appear in $C_p$($T$) curve just below 11 K. When the applied field exceeds 2 T, this weak anomaly transforms into a sharp and symmetric peak. The sharp nature of the peak suggests that the transition is first-order.  As this phenomenon occurs in the neighborhood of  strong AFM ordering transition, it is difficult to determine the exact position of the peak. We have determined the positions of the peaks by fitting a pair of Gaussian functions  plus a background to the data and found that the peak positions do not shift with field. Also, $C_p$ in the low temperature region  is gradually enhanced with increasing field  up to 5 T and then decreases. In the antiferromagnetic state, initially magnetic entropy (and hence $C_p$) increases with $H$ due to the increase of field-induced spin disordering  in one of the magnetic sublattices which is antiparallel to applied field. At high fields, however,  the volume fraction of FM phase increases significantly and as a result, magnetic entropy reduces. In several AFM systems where field-induced AFM-FM transition occurs, with increasing magnetic field, magnetic entropy initially increases up to a critical field and then decreases  \cite{mid1,mid2}. \\
\begin{figure}[h]
\includegraphics[width=0.7\textwidth]{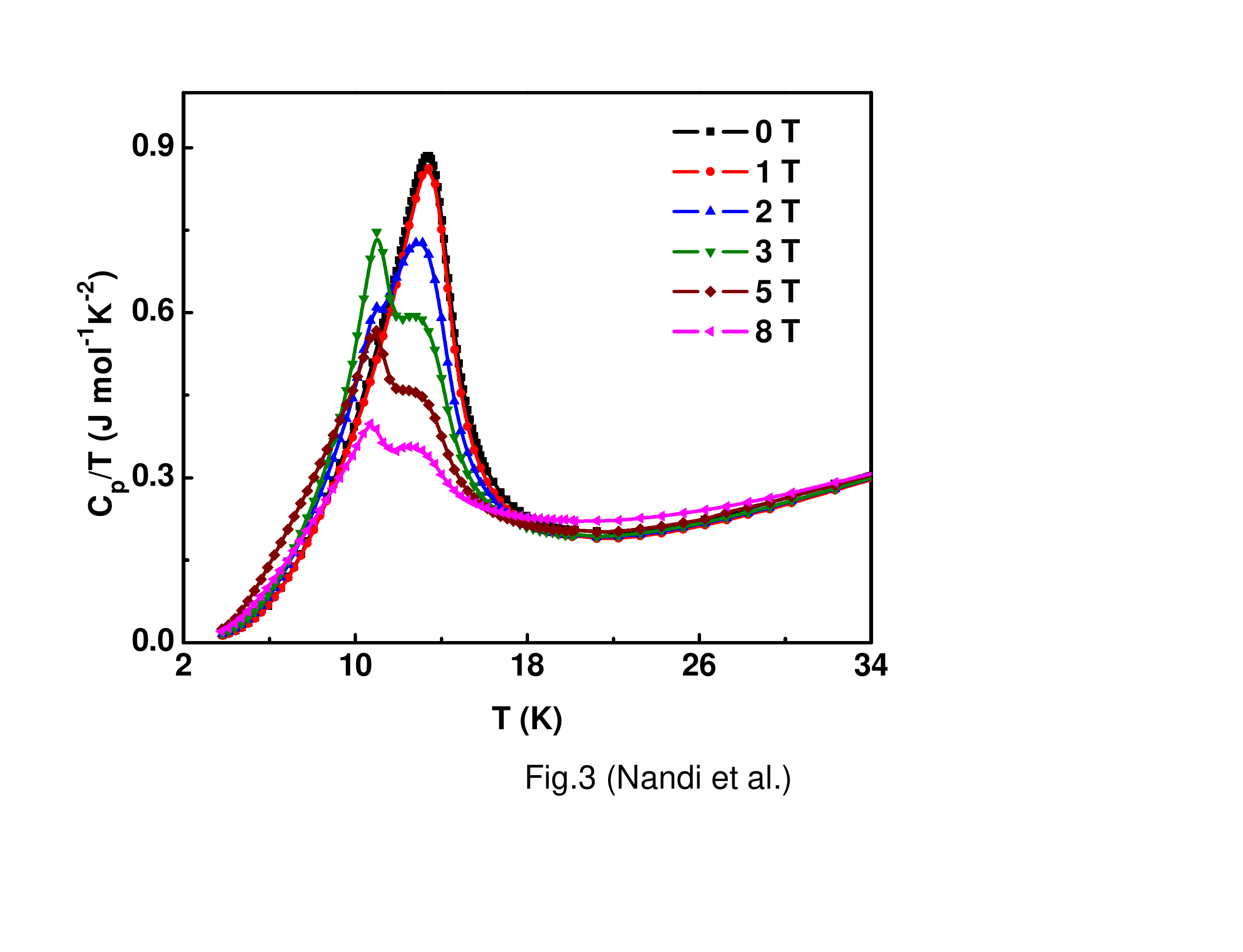}
\caption{ $C_p/T$ versus $T$ plots for $\alpha$-CoV$_2$O$_6$  for different fields in the vicinity of antiferromagnetic transition.}\label{fig3}
\end{figure}

We now compare the observed specific heat results of $\alpha$-CoV$_2$O$_6$ with that reported for several other frustrated magnetic systems.  The two-peak feature in the temperature dependence of the specific heat has been extensively investigated both in theory and experiments \cite{taba,ram,hir,anto,li,hond,yoshi}. Several ferrimagnets  and quasi-1D AFM spin chains exhibit field-induced peak  in $C_p$ due to the opening of an energy gap at low temperature \cite{li,hond,yoshi}. In these cases, however, the peak  is not very sharp and symmetric. Also, the peak shifts and  broadens with increasing field strength. In Dy$_2$Ti$_2$O$_7$ pyrochlore, the zero-field $C_p$($T$) exhibits a single peak  but a new sharp and symmetric peak appears at very low temperature  with  the application of magnetic field  \cite{ram,hir}. We have magnetically aligned the polycrystalline powder sample and done detailed $M$($H$) and $M$($T$) measurements to understand the field and temperature induced transitions. For this aligned sample, $M$($H$) at 5 K and $M$($T$) curves for three different fields are shown in the supporting information section (please see Figures S2 and S3 in the Supporting Information). Similar to randomly oriented powder sample, $M$($T$) shows  AFM to PM transition at 15 K for applied field below $H_{c1}$ but $M$($T$) exhibits a sharp peak at 11 K in the field range 1.5-3.3 T due to the transition from ferrimagnetic to PM state. Above 3.3 T where the system is in field-induced FM state, the nature of $M$($T$) curve is very different. With increasing temperature, $M$ decreases very slowly up to about 8 K and then decreases at a faster rate. This behavior of $M$($T$) is typical of a FM system in presence of high magnetic field. Though,  the transition region gets smeared, d$M$/d$T$ reveals a minimum at around 11 K. So it is clear that both ferrimagnetic-PM and FM-PM transitions occur almost at same temperature (11 K).  For this reason, we have not observed a third peak in $C_p$($T$) curve at high fields above 3.3 T due to the FM-PM transition. In polycrystalline samples, however, for some crystallites the effective field along the easy axis of magnetization direction can be significantly smaller than the applied field  due to their random orientations. Thus, when the applied field exceeds $H_{C1}$ or $H_{C2}$, these crystallites remain in the AFM or ferrimagnetic state and, as a result, different magnetically ordered phases coexist in the  sample. This observation is fully consistent with the reported low-temperature neutron diffraction results \cite{lener}.  Powder neutron diffraction studies have revealed the coexistence of different magnetically  ordered phases  in $\alpha$-CoV$_2$O$_6$ when the strength of applied magnetic field exceeds $H_{C1}$ or $H_{C2}$ \cite{lener}. For example, both AFM and ferrimagnetic phases coexist with volume fractions  0.54 and 0.46, respectively at 2.5 T. However,  both the phases decrease substantially  and FM phase appears at fields above $H_{C2}$.  At 5 T, the volume fractions of AFM, ferrimagnetic and FM phases are 0.20, 0.14 and 0.66, respectively.\\

In order to shed some more light on the nature and origin of the field-induced peak in $C_p$, we have studied the linear expansion of sample length [$\Delta L/L$] as  functions of $T$ and $H$. Figure 4(a) presents  $\Delta L$($T$)/$L_{4K}$ ($=$[$L$($T$)-$L_{4K}$]/$L_{4K}$)  plots up to 45 K for some selected fields, where $L_{4K}$ is the length of the sample at $T$$=$4 K in an applied field $H$. The inset of Fig. 4(a) shows $\Delta L$($T$)/$L_{4K}$ plot up to 300 K for $H$$=$0.  At zero field, though $\Delta L/L_{4K}$ decreases monotonically with decreasing $T$, it starts to decrease at a much faster rate as $T$ approaches $T_N$. In the PM state,  $\Delta L/L_{4K}$ is approximately linear in $T$ below 50 K for field up to 3 T but it develops a broad maximum close to $T_N$ and shows an upward curvature above $T_N$ for higher fields. Though, the anomaly at $T_N$ gets weakened with increasing field strength, it remains visible up to the highest applied field 9 T. At and above 1 T, $\Delta L$($T$)/$L_{4K}$ shows a step-like  decrease below a critical temperature $T_S$$=$8 K . This decrease in $\Delta L/L_{4K}$ at $T_S$ is  much sharper in nature than that at $T_N$.  To keep track on the field-evolution of the anomalies at $T_N$ and $T_S$,  we have determined $\alpha$($T$) at different fields from the slope of the curves in Fig. 4(a).  Figure 4(b) illustrates the temperature variation of the respective $\alpha$($T$). At zero field, $\alpha$($T$) is positive over the whole temperature range. Similar to $C_p$($T$), $\alpha$($T$) exhibits  a sharp $\lambda$-like peak close to $T_N$  and the peak shifts to lower temperature and broadens with increasing field. Over a narrow temperature range in the PM state, $\alpha$ becomes small and negative for applied field above 5 T. The field-induced peak in $\alpha$($T$) at $T_S$ is extremely sharp and symmetric and its position is almost insensitive to $H$. This peak is well separated from AFM transition and visible only below 5 T. In polycrystalline samples, the linear thermal expansion measured using dilatometer is just an average of the three crystallographic axis directions. Normally, the coefficient of linear thermal expansion  measured using dilatometer is 1/3 of the volume expansion coefficient \cite{souza}. We have compared the present results on thermal expansion with that reported from the temperature variation of lattice parameters determined using powder neutron diffraction \cite{mark,markk}. Indeed, the nature of $T$ dependence of the linear thermal expansion is quite similar to the volume expansion determined from powder neutron diffraction data.\\
\begin{figure}[th]
\includegraphics[width=0.425\textwidth]{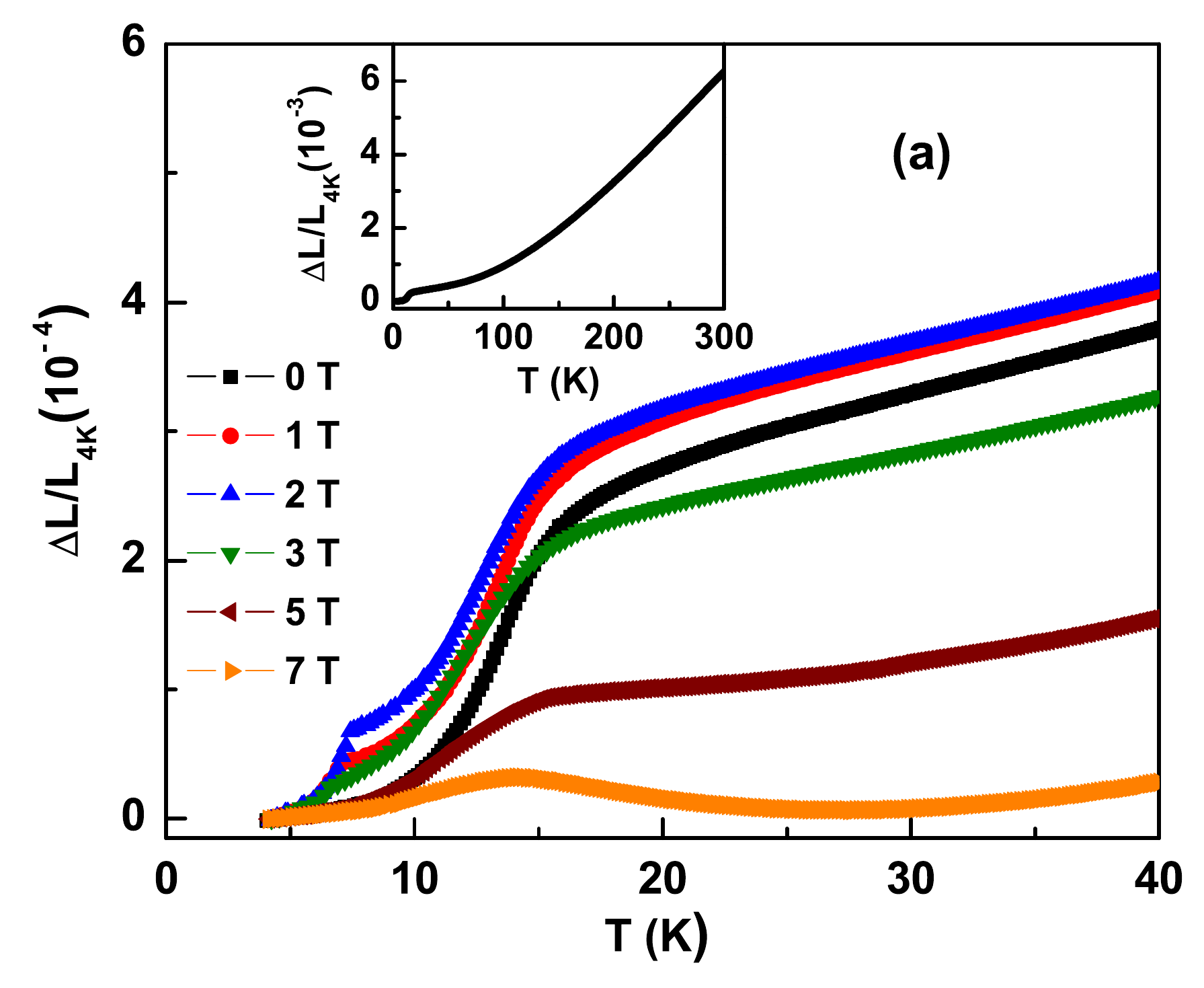}
\includegraphics[width=0.425\textwidth]{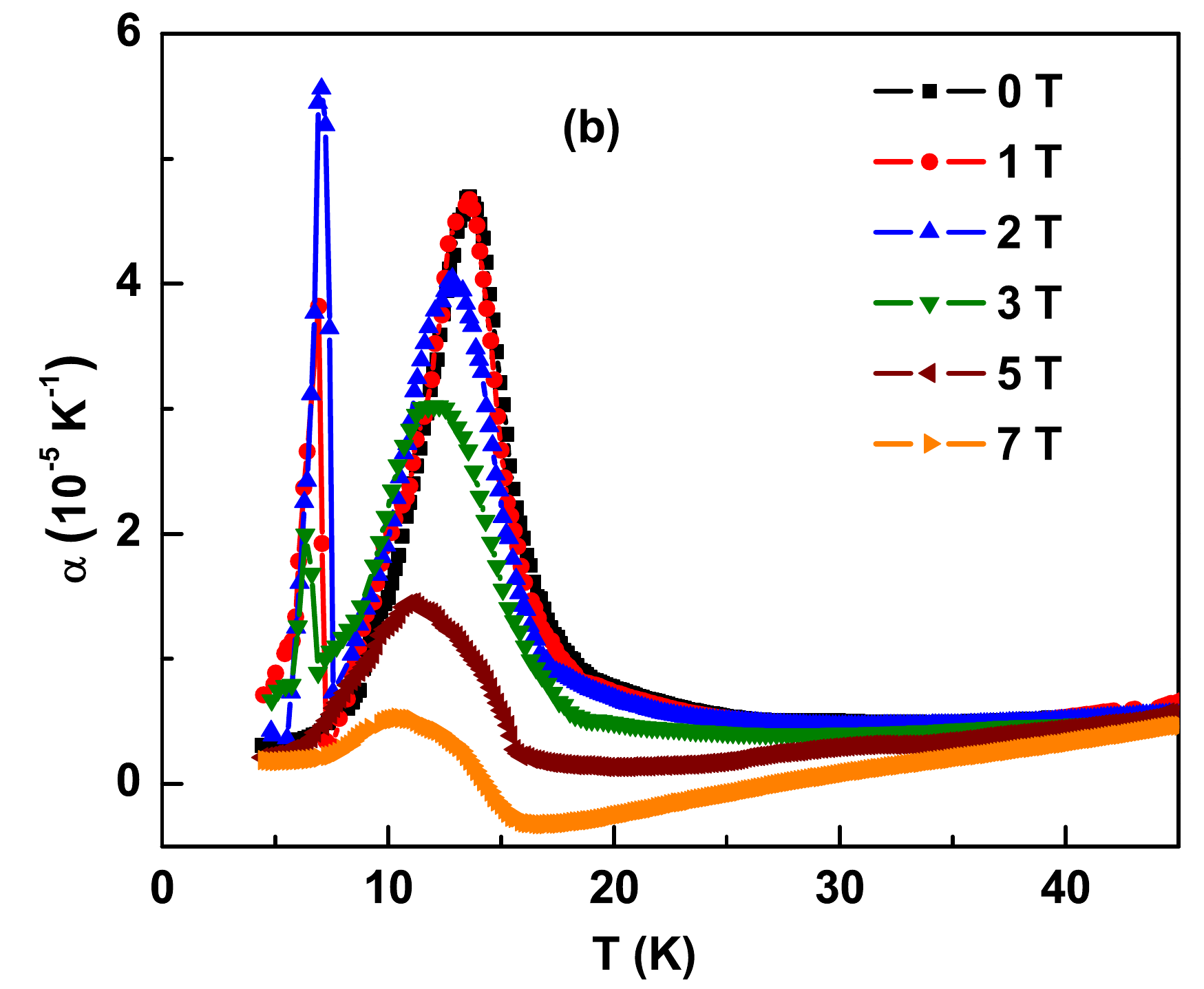}
\includegraphics[width=0.425\textwidth]{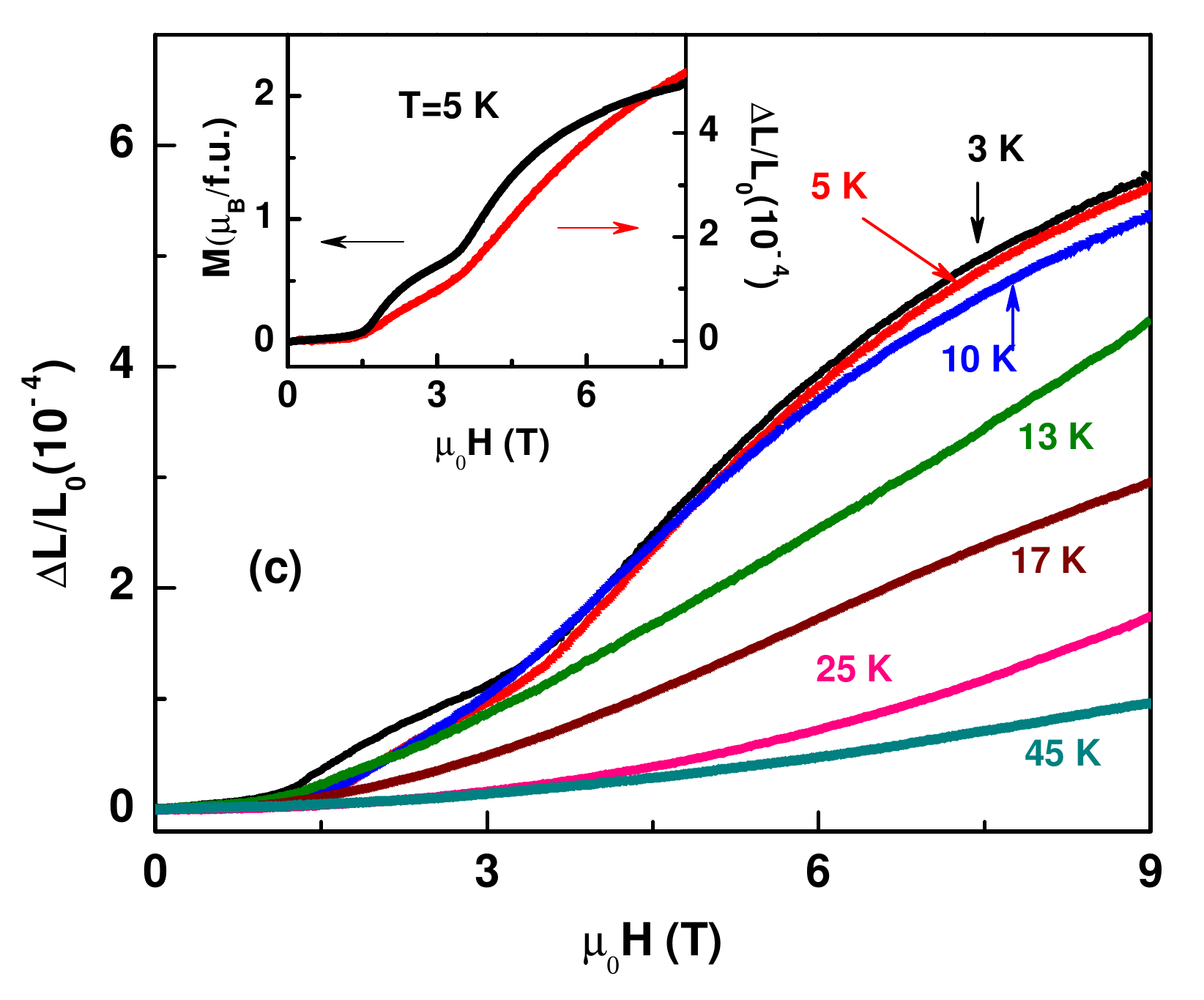}
\caption{(a) The temperature dependence of the relative length change ($\Delta L/L_{4K}$) for $\alpha$-CoV$_2$O$_6$ for different fields. Inset: $\Delta L($T$)/L_{4K}$ in zero field in the temperature range 4-300 K. (b) Plots of the thermal expansion coefficient $\alpha$($T$) for different fields. (c) Magnetostriction, $\Delta L(H)/L_0$,  at several temperatures both below and above $T_N$. Inset: the field dependence of magnetization and $\Delta L/L_0$ at 5 K.}\label{fig4}
\end{figure}
The magnetostriction, $\Delta L$($H$)/$L_0$$=$[$L$($H$)-$L_0$]/$L_0$, where  $L_0$ is the length of the sample  in absence of magnetic field, for some selected temperatures both below and above $T_N$ is depicted in Fig. 4(c). Figure shows that the field dependence of magnetostriction is very sensitive to  temperature.  At low temperatures ($T$$\leq$10 K), $\Delta L$/$L_0$ is  almost  independent of $H$ up to $H_{c1}$$\sim$1.5 T.  Above $H_{c1}$, $\Delta L$($H$)/$L_0$  increases rapidly  and exhibits a weak anomaly around $H_{c2}$. Thus,  $\Delta L$($H$)/$L_0$ exhibits anomaly at two critical fields that correspond to field-induced metamagnetic transitions in $M$($H$) curve   \cite{zhe,lene}. Similar to $M$($H$), these  anomalies are clearly visible at low temperatures  but they become weak  with increasing $T$ and disappear  above $T$$\simeq$10 K \cite{zhe,lene}. The disappearance of these anomalies  in both $M$($H$) and $\Delta L$($H$)/$L_0$ curves above 10 K  may be related to the observed field-induced first order transition  in $C_P$($T$). At low temperature, $\Delta L$($H$)/$L_0$ approximately mimics the nature of $M$($H$) curve  [inset of Fig. 4(c)]. $\Delta L$($H$)/$L_0$ develops a downward curvature with the increase of $H$ but it does not show any tendency of saturation up to the maximum applied field.  The absence of saturation in $\Delta L$($H$)/$L_0$ is consistent with the nature of $M$($H$) curve.  $M$ in single crystals or magnetically aligned sample with field along the $c$ axis saturates above 4 T while $M$  at 7 T is just 50$\%$  of the expected value in magnetically random polycrystalline sample \cite{zhe,lener}. The downward curvature in $\Delta L$($H$)/$L_0$ decreases with increasing $T$ and  $\Delta L$($H$)/$L_0$ becomes almost linear for temperatures close to or slightly above $T_N$. Well above $T_N$, $\Delta L$($H$)/$L_0$ decreases rapidly with increasing $T$  and develops a weak upward curvature.  The magnetostriction effect is very small above 50 K.  The unusually high value of the magnetostriction below $T_N$ is comparable to that reported for the frustrated spinels  ZnCr$_2$Se$_4$ and CdCr$_2$O$_4$ where the spin-lattice coupling is believed to be quite strong \cite{loidl,ueda}. In these spinels, however, the nature of  $H$ dependence of $\Delta L$/$L_0$  is  different from that for the $\alpha$-CoV$_2$O$_6$ system. Furthermore, ZnCr$_2$Se$_4$ exhibits a strong nearly constant negative thermal expansion over a wide range of $T$ just above $T_N$ which has been attributed to the geometrical frustration of the lattice degrees of freedom \cite{loidl}. In $\alpha$-CoV$_2$O$_6$, however, $\alpha$($T$) is small and negative only at high fields. The appearance of the peak in  $C_p$($T$) and $\alpha$($T$) with applied field and  the large positive magnetostriction effect below $T_N$ suggest that the field-induced magnetostructural transition occurs due to the strong spin-lattice coupling and the field-induced FM state occupy more space than the AFM state.  We believe that the values of thermal expansion and magnetostriction can be significantly larger in single crystal. \\

\section{Conclusions}
In conclusion, $C_p$($T$) and  $\alpha$($T$) of the quasi-1D antiferromagnetic spin chain $\alpha$-CoV$_2$O$_6$ exhibit a sharp $\lambda$-like peak around $T_N$. The magnetic entropy calculated from $C_p$($T$) data shows that  the magnetic fluctuations persist well above $T_N$ but  the system behaves like a spin-1/2 Ising chain below 20 K.  In the AFM state, both $C_p$ and  $\alpha$ exhibit strong $T$ and $H$ dependence. When the applied field exceeds $H_{c1}$,  a sharp and symmetric peak emerges well below $T_N$ due to the ferromagnetic/ferrimagnetic-paramagnetic transitions. The sharp nature of the peak indicates that the transition is first order. The huge magnetostriction effect below $T_N$ and its strong $H$ dependence  along with the field-induced spin structure transition suggest that the spin-lattice coupling in this system is  quite strong. The measurements of $C_p$ and $\Delta L$/$L$ on single crystal with $H$  along different crystallographic axes may reveal interesting physical properties for $\alpha$-CoV$_2$O$_6$.\\

\section{Associated Content}
Supporting Information\\
In the temperature range 50-222 K, $C_p$ of $\alpha$-CoV$_2$O$_6$ has been fitted with the Einstein model of lattice specific heat [Figure S1]. The temperature and magnetic field dependence of magnetization have been measured for the magnetically aligned polycrystalline powder sample [Figure S2 and S3]. This material is available free of charge via the Internet at http://pubs.acs.org.\\

\newpage
\appendix
\begin{center}
\section{Supporting Information}
\end{center}
The temperature dependence total specific heat ($C_p$) is shown in Fig. S1. In the range 50-222 K, $C_p$ is fitted well by the Einstein model of lattice specific heat. In order to extract the lattice contribution to $C_p$ in the paramagnetic (PM) as well as in the antiferromagnetic (AFM) states, the Einstein fitting curve has been extrapolated down to lowest measured temperature (2 K).

\begin{figure}
\renewcommand{\thefigure}{S{1}}
\includegraphics[width=0.7\textwidth]{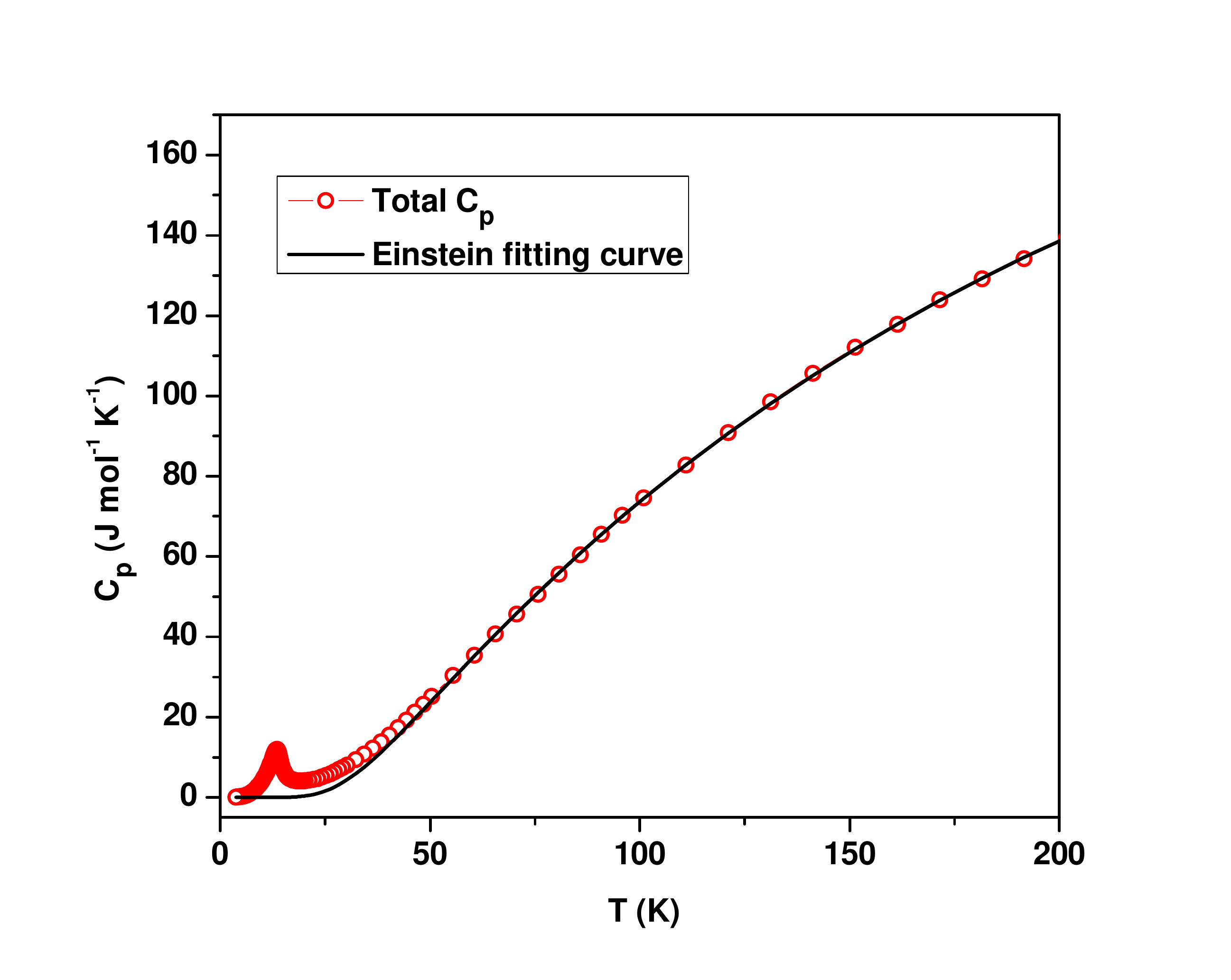}
\caption{The fitting of the specific heat ($C_p$) data with the Einstein model of lattice specific heat. The black line represents the Einstein fitting curve and the red symbol represents the total specific heat.}\label{figS1}
\end{figure}

We have aligned the powder sample of $\alpha$-CoV$_2$O$_6$ in a polymeric glue by applying 7 T magnetic field at room temperature which freezes the particle orientation. The field dependence of magnetization (M) at 5 K is shown in Fig. S2. It is clear from the figure that M(H) exhibits sharp field-induced transitions at two critical fields ($H_{c1}$ and $H_{c2}$) and 1/3 magnetization plateau between $H_{c1}$and $H_{c2}$, and the saturation value of M in the ferromagnetic (FM) state (above $H_{c2}$) is about 4.6 $\mu_B$/Co.

\begin{figure}
\renewcommand{\thefigure}{S{2}}
\includegraphics[width=0.7\textwidth]{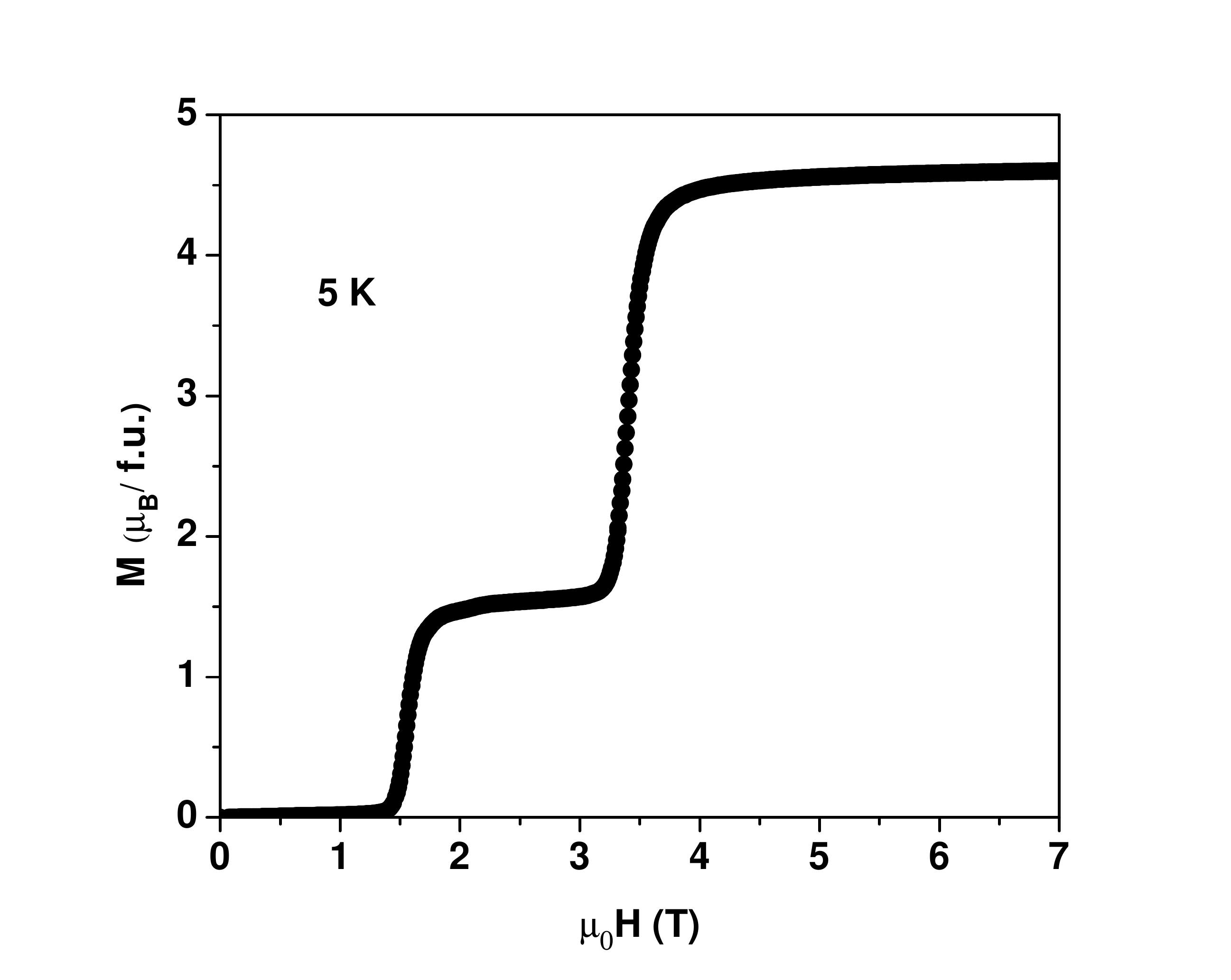}
\caption{The field dependence of magnetization of aligned $\alpha$-CoV$_2$O$_6$ sample at 5 K.}\label{figS2}
\end{figure}
Figure S3 shows representative M(T) curves at three different magnetic fields of aligned polycrystalline $\alpha$-CoV$_2$O$_6$. At 0.1 T, an AFM-PM transition occurs around 15 K (upper panel). For $H_{c1}$<H<$H_{c2}$, the peak due to AFM-PM transition disappears but a new peak appears at lower temperature (11 K) due to ferrimagnetic-PM transition (middle panel). However, above $H_{c2}$, the nature of M(T) curve is typical of a FM system in high field with FM-PM transition $\sim$11 K (lower panel). This study shows that $\alpha$-CoV$_2$O$_6$ undergoes ferrimagnetic-PM and FM-PM transitions at same temperature (11 K).

\begin{figure}
\renewcommand{\thefigure}{S{3}}
\includegraphics[width=0.7\textwidth]{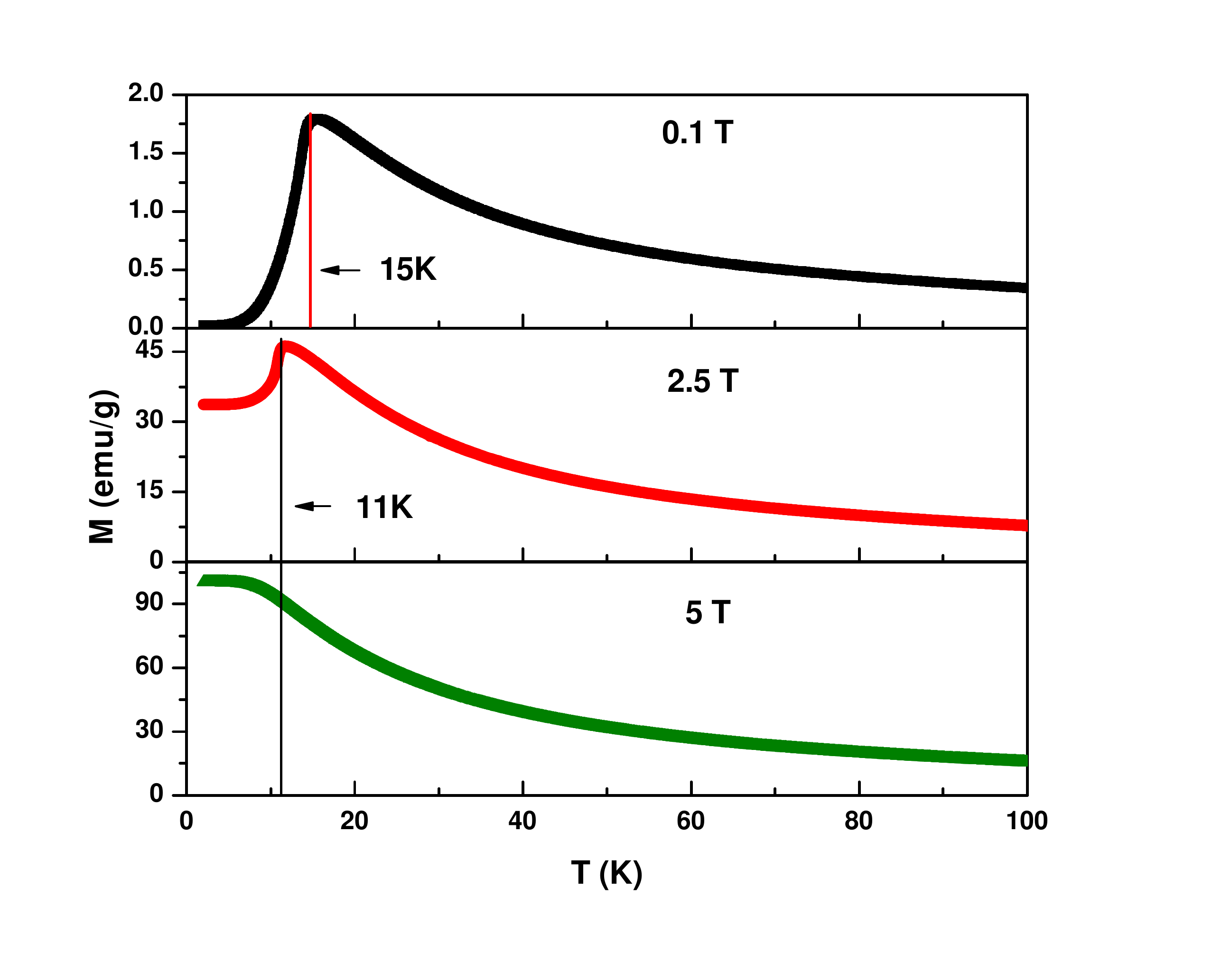}
\caption{Three representative M(T) curves for the  magnetically aligned $\alpha$-CoV$_2$O$_6$ sample.}\label{figS3}
\end{figure}

\end{document}